\documentclass[journal=jctcce,manuscript=article]{achemso}
\usepackage[utf8]{inputenc}
\setkeys{acs}{maxauthors=500}
\usepackage{amsmath,amssymb}

\usepackage{bm}

\renewcommand{\d}{\,\mathrm{d}}
\renewcommand*{\vec}[1]{\boldsymbol{#1}}

\title{Accurate transferable polarization model derived from the monomer electron density}
\author{Ruben Goeminne}
\affiliation[Ghent University]{Center for Molecular Modeling (CMM), Ghent University,\\ Technologiepark 46, 9052 Zwijnaarde, Belgium}
\author{Toon Verstraelen}
\affiliation[Ghent University]{Center for Molecular Modeling (CMM), Ghent University,\\ Technologiepark 46, 9052 Zwijnaarde, Belgium}
\email{Toon.Verstraelen@UGent.be}

\begin{document}

	\maketitle


	\vfill


	\vspace{1cm}

	\newpage

	\begin{abstract}

	Force field have for decades proven to be an indispensable tool for molecular simulations which are out of reach for \textit{ab-initio} methods. Recent efforts to improve the accuracy of these simulations have focused on the inclusion of many-body interactions in force fields. In this regard, we propose a transferable inducible dipole model which requires only the monomer electron density as input, without the need for atom type specific parameters. Slater dipoles are introduced, the widths of which are derived from the \textit{ab-initio} monomer density. An additional exchange-repulsion interaction is introduced in our model, originating from the overlap of the delocalized dipoles with other dipoles and the ground state electron density. This interaction has previously been neglected in point dipole models, as the lack of spatial extent of the dipoles prevents the inclusion of an overlap term. The inclusion of this interaction is shown to significantly improve the prediction of three-body energies. Our model is incorporated in a previously proposed non-covalent force field and is benchmarked on interaction energies of dimers contained in the hsg and hbc6 datasets. Furthermore, we demonstrate the transferability of our model to the condensed phase of water, and to the interaction of CO$_2$ and H$_2$O molecules with the ZIF-8 metal-organic framework. The inherent transferability of our model makes it widely applicable to systems like the aforementioned metal-organic frameworks, where no specifically fitted parameters for polarization models are available in the literature.
	\end{abstract}

	\newpage

	\section{Introduction}

	The importance of non-additive effects in molecular simulations has long been established. Most efforts to include these many-body effects have focused on modeling the polarization component. This interaction is known to be important for the solvation free energy of salt ions and amines\cite{Wu2010,Jiao2006,Meng1996}, cation-$\pi$ interactions\cite{Caldwell1995}, the modeling of polarizable organic compounds\cite{Soetens1999,Wang2011a,Ren2011,Masella2003,Kaminski2002,Cieplak2009} and the various anomalous properties of water.\cite{Laury2015,Ren2003,Chen2000,Dang1998,Wang2012a} Models to include electronic polarization can generally be divided into the Drude oscillator,\cite{Lamoureux2003,Yu2003} fluctuating charge\cite{Rick1994} and inducible dipole models.\cite{Caldwell1990} The Drude oscillator or charge-on-spring model introduces an auxiliary charged particle which is attached to the polarizable center by a parabolic restraint. In the fluctuating charge model, the magnitude of the charge on the polarizable center itself is allowed to fluctuate. The inducible dipole model, on the other hand, does not alter the monopoles but introduces a basis of atomic dipole response functions at each atomic site. The dipole response has previously been modeled by point dipoles, in which case damping functions are required to avoid the polarization catastrophe\cite{Thole1981,Wang2012a}, or Gaussian dipoles, for which widths of Gaussian charge distributions are required. Furthermore, the parametrization of polarizabilities usually requires the introduction of a set of atom types, dependent on both the element and its chemical environment.\cite{Elking2007,Donchev2006} The atomic polarizabilities are then determined by fitting to the experimental or \textit{ab-initio} polarizability tensors for a set of molecules or functional groups which contain the atom type of interest, or by probing molecules with point charges or external electric fields.\cite{Elking2007,Thole1981,Masia2005} However, these methods limit the model's applicability to atoms in molecules with similar chemical environments as in the training set.

	This work is based on the inducible dipole model. Compared to models in the literature, we aim foremost at a fully transferable polarization model. It is transferable in a different sense than previous inducible dipole models. Usually, the term refers to the ability of a model to generalize to molecules outside the training set, which are however composed of atom types included in the training set. In contrast, our model makes use of the ground state electron density, obtained from a single \textit{ab-initio} calculation of the monomer(s) of interest. An interaction parameter is introduced, and is shown to transfer well to molecules outside the training set, making the model easily applicable to new systems without the need to define for atom types.

	We introduce Slater dipoles as response functions, and include an additional exchange interaction in our model, together with the previously used classical electrostatic interaction. The exchange is modeled as a proportionality with the overlap between electron densities, based on previous observations of this proportionality. \cite{Kim1981,Misquitta2016,Stone2013,Vandenbrande2017} After the inducible dipole model is introduced in Section \ref{model}, and Slater widths for each chemical element are determined in Section \ref{widths}, we validate  our model by its performance in the prediction of molecular polarizabilities in Section \ref{molpol} and a large number of three-body interaction energies in Section \ref{threebody}. In this way, we can directly compare the non-additive component of our model to \textit{ab-initio} three-body energies, ensuring our model exhibits the correct many-body behavior. In Section \ref{p-medff}, the inducible dipole model is included in the non-covalent monomer electron density force field (MEDFF), which decomposes the interaction energy in four interaction terms in accordance with SAPT.\cite{Jeziorski1994a} Similar to the newly proposed induction model, MEDFF only requires the monomer electron density as input. It comprises 3 interaction parameters which were fit to a set of dimer interaction energies, one of which is replaced by including the new inducible dipole model with its single interaction parameter. The resulting polarizable monomer electron density force field (PMEDFF) is benchmarked on dimer interaction energies in Section \ref{dimer}. Lastly, the transferability of our force field to the condensed phase (Section \ref{liquidwater}) and insertion energies of CO$_2$ and H$_2$O in the ZIF-8 metal-organic framework (Section \ref{mof}) is demonstrated.

	\newpage
	\section{Inducible dipole model}
	\label{model}

	A physical inducible dipole model is built up from essentially three components. The first is a determination of the atomic polarizabilities. The second is a representation of the ground state electron density and the functional form of the atomic dipole response functions. The last component is a model for the interactions of the induced dipoles with the ground state and for the interactions between the induced dipoles. These interactions have previously been approximated as purely electrostatic contributions. As the proposed polarization model differs in its three components from frequently-used inducible dipole models in the literature, these separate components are introduced first.

	\subsection{Atomic polarizabilities}
	\label{polarizabilities}

	Atomic polarizabilities are usually derived by defining a limited set of atom types, after which the respective atomic polarizabilities are fit to reproduce the experimental\cite{Wang2011} or \textit{ab-initio}\cite{Elking2007} derived molecular polarizability tensor. Our approach differs from this methodology, as our intent is to derive all model parameters except a single interaction parameter from an \textit{ab-initio} calculation of the molecule of interest. Therefore, we start from the free atom polarizabilities obtained from linear response time-dependent density functional theory.\cite{Chu2004} Subsequently, to account for the chemical environment, the isotropic polarizability of each atom $\alpha_{i, \text{free}}$ is rescaled proportionally to the effective volume ($V_\text{aim}$) of the atom-in-molecule

	\begin{equation}
	\alpha_i=\frac{V_{\text{aim}}}{V_\text{free}}\;\alpha_{i, \text{free}}=\frac{\int r^3\rho_i(\vec{r})\d\vec{r}}{\int r^3\rho_{i,\text{free}}(\vec{r})\d\vec{r}}\;\alpha_{i, \text{free}}
	\end{equation}

	with $V_\text{free}$ the effective volume of the free atom in vacuum. This rescaling of polarizabilities was proposed previously in the exchange-hole dipole moment (XDM) and Tkatchenko-Scheffler dispersion models.\cite{Becke2006,Tkatchenko2009}
	For consistency with MEDFF, all ground state densities were calculated at the spin-polarized B3LYP/aug-cc-pVTZ level of theory. Similarly, for consistency with the Slater dipoles derived in the following Section, the minimal basis iterative stockholder (MBIS) scheme was used to partition the molecular density into atomic fragments.\cite{Verstraelen2016}

	\subsection{Slater dipole response functions}
	\label{slaters}

	Previously, both point charges and Gaussian charge distributions have been used to represent the ground state electron density, together with point dipoles and Gaussian dipoles as response functions. However, as the true electron density tails off exponentially, atom-centered Slater functions are used in this work to represent the ground state.\cite{Ahlrichs1972,OConnor1973,Hoffmann-Ostenhof1977} To this end, the MBIS scheme \cite{Verstraelen2016} is used to partition the \textit{ab-initio} molecular density into atom-centered core charges $q^\text{c}_A$ and valence 1s Slater functions of the form:

	\begin{equation}
	\rho^\text{1s}_{A}(\vec{r})=\frac{N_A}{8\pi{\sigma_{A,\text{s}}}^3}\exp\left(-\frac{\left|\vec{r}-\vec{R}_A\right|}{\sigma_{A,\text{s}}}\right)+q^\text{c}_A\, \delta(\vec{r}-\vec{R}_A)
	\end{equation}

	with $N_A$ the population and $\sigma_{A,\text{s}}$ the width of the distribution, both of which are fitted to the \textit{ab-initio} density by minimizing the Kullback-Leibler divergence.
	This partitioning has been show to accurately reproduce dimer electrostatic interactions, while being robust with respect to small changes in the electronic structure calculation from which it is derived\cite{Verstraelen2016}. A Slater dipole can now be constructed as the gradient of a normalized 1s Slater function. However, as this function has no well defined limit towards the $\vec{R}_A$, we alternatively use a normalized 1s+2s function:

	\begin{align}
	\vec{\rho}^\text{1p}_{A}(\vec{r})&=\vec{\nabla}_A\,\rho^\text{1s+2s}_{A}\\
	&=\vec{\nabla}_A\left[\frac{1}{32\pi{\sigma_{A,\text{p}}}^3}\exp\left(-\frac{\left|\vec{r}-\vec{R}_A\right|}{\sigma_{A,\text{p}}}\right)\left(1+\frac{\left|\vec{r}-\vec{R}_A\right|}{\sigma_{A,\text{p}}}\right)\right]\\
	&=\frac{1}{32\pi{\sigma_{A,\text{p}}}^5}\exp\left(-\frac{\left|\vec{r}-\vec{R}_A\right|}{\sigma_{A,\text{p}}}\right)\left(\vec{r}-\vec{R}_A\right)
	\end{align}

	This Slater dipole can be interpreted as the density difference between an unperturbed ground state electron density and a density perturbed by an electric dipole field. From this interpretation, the dipole width $\sigma_{A,\text{p,free}}$ for each free atom can be determined by fitting a Slater dipole to the normalized \textit{ab-initio} density difference of a free atom. More details are provided in Section \ref{widths}. To account for the chemical environment of each atom in a molecule or solid, every free atom dipole width $\sigma_{A,\text{p,free}}$ is subsequently scaled with the cube root of the ratio of its effective volume in a molecule compared to its effective volume in vacuum\cite{Tkatchenko2009}, similarly to the rescaling used for the polarizabilities in Section \ref{polarizabilities}:

	\begin{equation}
		\sigma_{A,\text{p}}=\left(\frac{\int r^3\rho_A(\vec{r})\d\vec{r}}{\int r^3\rho_{A,\text{free}}(\vec{r})\d\vec{r}}\right)^{1/3}\sigma_{A,\text{p,free}}
	\label{rescale}
	\end{equation}

	\subsection{Interaction model}
	\label{interaction}
	Similarly to previous inducible dipole models,\cite{Cieplak2001,Masia2005,Rasmussen2007,Kaminski2002,Xie2007,Elking2007} the induced dipole on each atom $\vec{d}_i$ can be determined by solving

	\begin{equation}
	\vec{d}_i=\alpha_i\left(\vec{E}^\text{sp}_i - \sum_{j\neq i}\vec{T}^\text{pp}_{ij}\cdot \vec{d}_j\right)
	\label{pff_eq}
	\end{equation}

	with $\alpha_i$ the isotropic polarizability of atom $i$, $\vec{E}^\text{sp}_i$ the field generated at site $i$ by the core and valence charges of the surrounding atoms, and $\vec{T}^\text{pp}_{ij}$ the interaction tensor between a dipole at site $i$ with a dipole at site $j$. The second term $- \sum_{j\neq i}\vec{T}^\text{pp}_{ij}\cdot \vec{d}_j$ consequently represents the field generated at site $i$ due to all other induced dipoles $\vec{d}_j$. Both the monopole-dipole ($\vec{E}^\text{sp}_i$) and dipole-dipole ($\vec{T}^\text{pp}_{ij}$) terms represent interactions between charge distributions which in previous polarizable force fields have been approximated as purely electrostatic:

	\begin{equation}
	\vec{E}^\text{sp}_i=\sum_{j\neq i} \iint \frac{\vec{\rho}^\text{1p}_i(\vec{r})\rho^\text{1s}_j(\vec{r}')}{\left|\vec{r}-\vec{r}'\right|}\d\vec{r}\d\vec{r}'
	\end{equation}

	\begin{equation}
	\vec{T}^\text{pp}_{ij}=\iint \frac{\vec{\rho}^\text{1p}_i(\vec{r})\vec{\rho}^\text{1p}_j(\vec{r}')}{\left|\vec{r}-\vec{r}'\right|}\d\vec{r}\d\vec{r}'
	\end{equation}

	This approximation only captures the classical electrostatic interaction, and does not take into account the quantum mechanical exchange-repulsion interaction due to the antisymmetry constraint of the wave function with respect to the exchange of two electrons.\cite{Echenique2007} The need for an additional interaction which induces the dipoles has been noted before, with the magnitude of induced dipoles being consistently underestimated at intermediate distances.\cite{Masia2005,Vandenbrande2017} The exchange-repulsion interaction has previously been neglected due to the fact that it cannot be represented as a functional of the interacting electron densities. However, an approximate proportionality between this interaction and the overlap of the electron densities has been observed previously.\cite{Kim1981,Misquitta2016,Stone2013,Vandenbrande2017} Recently, the proportionality factor was determined by fitting to SAPT2+3 exchange-repulsion interaction energies of the dispersion dominated dimers in the S66x8 set, and is on average equal to 8.13 a.u. \cite{Rezac2011,Jeziorski1994a,Vandenbrande2017}. This overlap model reproduces SAPT2+3 exchange-repulsion energies with a root-mean squared error of 1.22 kcal/mol. This additional interaction is now included in our polarization model:

	\begin{equation}
	\vec{E}^\text{sp}_i=\sum_{j\neq i}\left[ \iint \frac{\vec{\rho}^\text{1p}_i(\vec{r})\rho^\text{1s}_j(\vec{r}')}{\left|\vec{r}-\vec{r}'\right|}\d\vec{r}\d\vec{r'}+U_\text{exch-ind}\int\vec{\rho}^\text{1p}_i(\vec{r})\rho^\text{1s}_j(\vec{r})\d\vec{r}\right]
	\label{field}
	\end{equation}

	\begin{equation}
	\vec{T}^\text{pp}_{ij}=\iint \frac{\vec{\rho}^\text{1p}_i(\vec{r})\vec{\rho}^\text{1p}_j(\vec{r}')}{\left|\vec{r}-\vec{r}'\right|}\d\vec{r}\d\vec{r}'+U_\text{exch-ind}\int\vec{\rho}^\text{1p}_i(\vec{r})\vec{\rho}^\text{1p}_j(\vec{r})\d\vec{r}
	\label{pff}
	\end{equation}

	where $U_\text{exch-ind}$ represents the proportionality between the overlap in electron density and the exchange-repulsion energy. In this polarization model, Slater dipoles can thus be induced to minimize both the electrostatic and exchange-repulsion energy between the ground state electron densities of molecules. We chose to let the sum in Eq. \ref{field} only run over atoms which do not belong to the same monomer as atom $i$, as the intramolecular polarization effects are already mostly captured in the \textit{ab-initio} monomer density used as input.

	\section{Results and Discussion}

	\subsection{Free atom dipole widths}
	\label{widths}

	Before the new inducible dipole model proposed in this work is benchmarked, the free atom dipole widths are determined for the following set of elements: H, C, N, O, F, Mg, Al, P, S, Cl and Zn. The response in electron density was determined by applying a dipole field of 0.0001 a.u.\ along the $z$-axis to the free atoms. This field strength is small enough to prevent higher order effects, but large enough to prevent numerical instabilities. Electron densities of both the perturbed and unperturbed free atoms were calculated with the CCSD method\cite{Purvis1982} and the aug-cc-pV5Z basis set\cite{Peterson1994} using Gaussian09\cite{g09} for all atoms except for zinc. For this element, scalar relativistic effects were included with the exact-two-component (X2C) method\cite{Verma2016} at the CCSD/ANO-RCC-VQZP level of theory\cite{Pritchard2019} in Psi4\cite{Parrish2017}. The large size of these basis sets was chosen to obtain accurate density tails.

\begin{figure}[t]
		\centering
		\includegraphics[width=\textwidth]{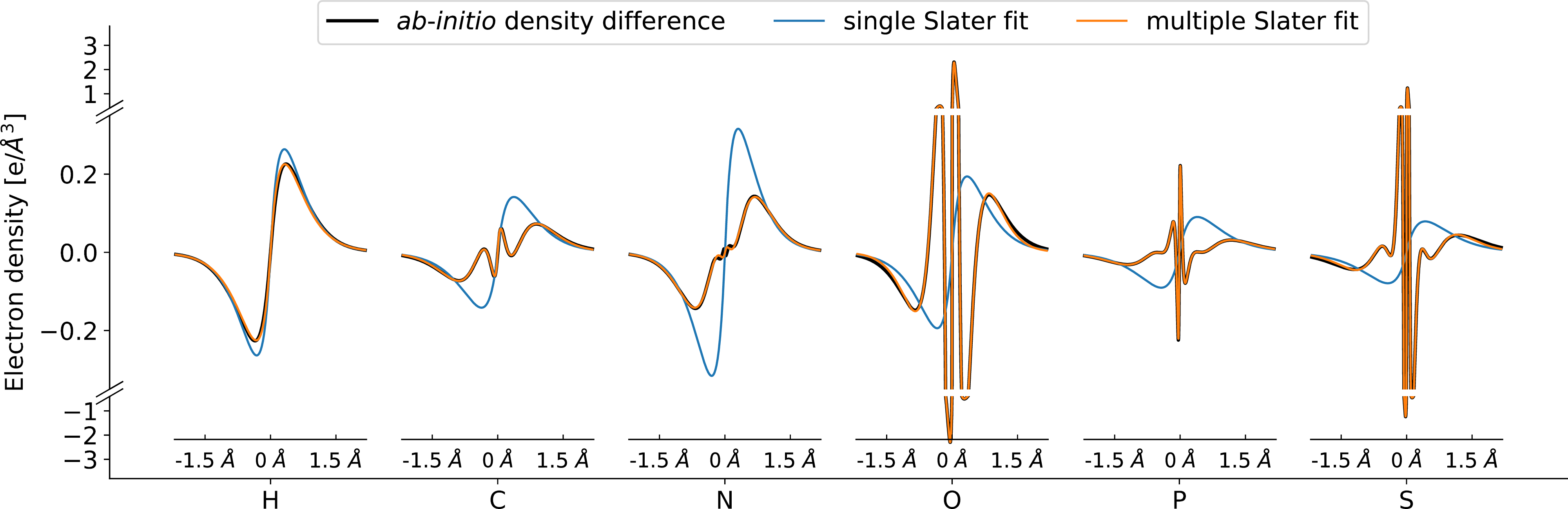}
		\caption{\textit{Ab-initio} density difference between the unperturbed free atoms and atoms in a dipole field along the $z$-axis (in black), the Slater function fit to the \textit{ab-initio} expectation value of $z^3$ (in blue), and the least squares fit of multiple Slater functions to the \textit{ab-initio} density (in orange).}
		\label{slaterwidths}
	\end{figure}

	The resulting density differences are shown in Figure \ref{slaterwidths} for H, C, N, O, P and S. As large fluctuations of the density close to the nucleus are present, a least squares fit of a single Slater function yields unreliable results. Therefore, the Slater widths were determined by fitting the third moment $\langle z^3\rangle$ of the Slater function to that of the \textit{ab-inito} density difference. More details are provided in the Supporting information. As seen from Figure \ref{slaterwidths}, this single Slater function per atom doesn't capture all fluctuations of the density, but is in reasonable agreement far from the nucleus. To investigate whether these fluctuations cause significant deviations in the energy of our polarization model, an additional fitting of multiple Slater functions per atom is also determined. In this case, a least squares fit was used, as it is desired to fit to the exact density fluctuations. Both the amplitudes and Slater widths were fit, and the lowest number of Slater functions able to capture all density fluctuations is used. The result are shown in orange in Figure \ref{slaterwidths}. For these elements, a very accurate fit to the \textit{ab-initio} density difference is possible using at most 5 Slater functions. Note that the widths of the derived atomic Slater functions in a molecule are subsequently rescaled proportional to the cube root of the effective volume according to Equation \ref{rescale} in order to account for their chemical environment.

\subsection{Reproduction of the molecular polarizability}
\label{molpol}

	\begin{figure}[t]
		\centering
		\includegraphics[width=0.9\textwidth]{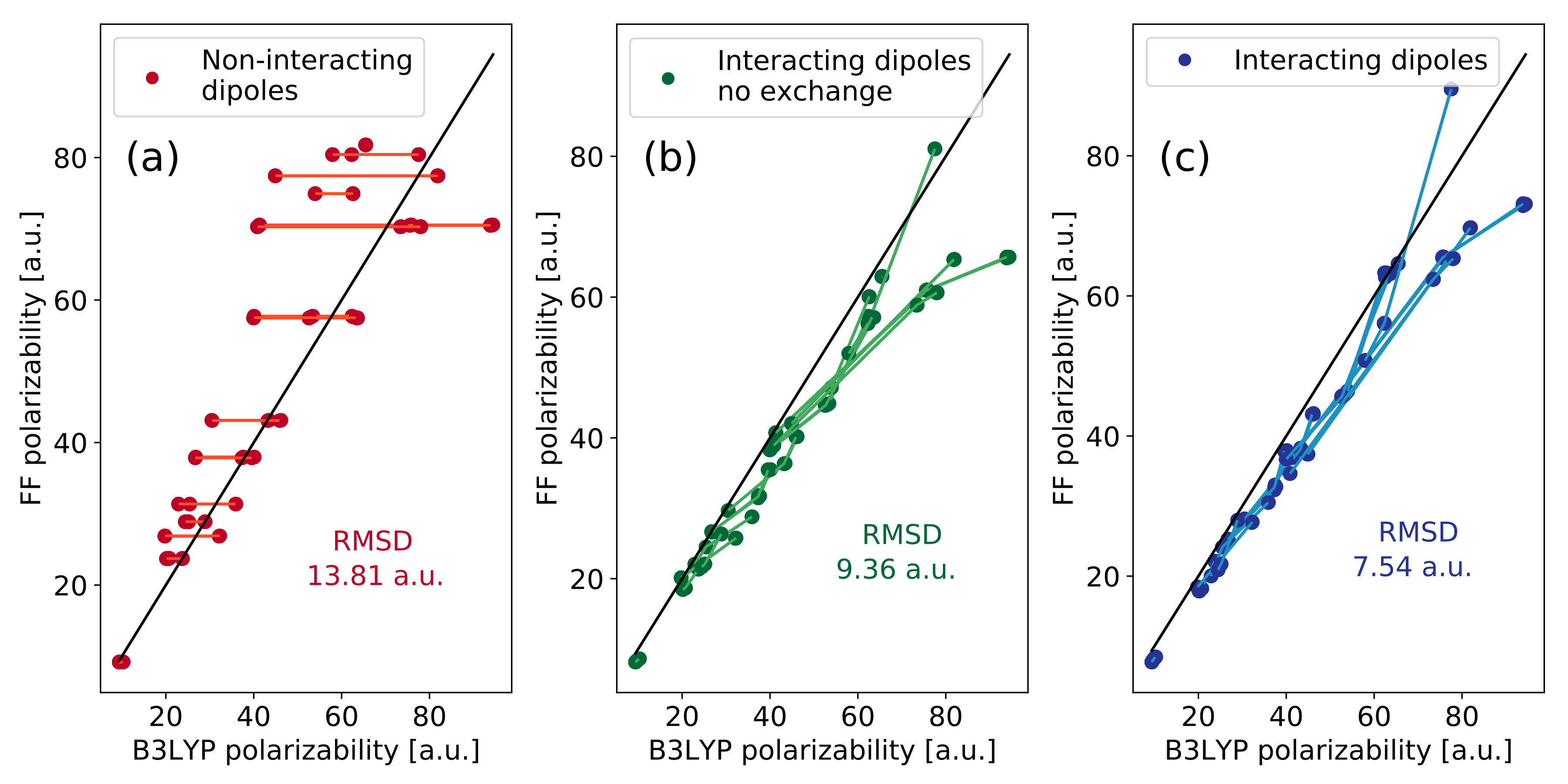}
		\caption{Eigenvalue parity plot of the molecular polarizabilities of monomers extracted from the S66 set. The eigenvalues, computed with (a) non-interacting dipoles, (b) interacting dipoles without the exchange interaction and (c) interacting dipoles with the exchange interaction are compared with the \textit{ab-initio} eigenvalues. The three eigenvalues for each monomer are connected to visualize the anisotropy of the polarizability tensor.}
		\label{molecular_pol}
	\end{figure}

	A first validation of the polarizable force field model developed in this work is the reproduction of the molecular polarizability tensor. By applying a uniform dipole field $\vec{E}^\text{sp}_i$ in equation \ref{pff_eq}, the atomic dipoles are calculated and summed together to equal the induced molecular dipole. If the dipole-dipole interaction tensor $\vec{T}^\text{pp}_{ij}$ is neglected, the molecular polarizability is simply equal to the sum of the atomic polarizabilities. Therefore, as we make use of isotropic atomic polarizabilities, the molecular polarizability is isotropic in this case. In contrast, by including $\vec{T}^\text{pp}_{ij}$ (defined in Equation \ref{pff}), a coupling is introduced between atomic dipoles along chemical bonds. This coupling gives rise to anisotropic contributions in the molecular polarizability tensor. Equation \ref{pff} can therefore be validated by the degree to which it gives rise to the correct molecular anisotropy.

	To perform this validation, the molecular polarizabilities of all monomer in the S66 set\cite{Rezac2011} were calculated at the B3LYP/aug-cc-pVTZ level of theory. First, the eigenvalues of the molecular polarizability tensors in the non-interacting dipole model ($\vec{T}^\text{pp}_{ij}=\vec{0}$) are shown in Figure \ref{molecular_pol}(a). For each monomer, the 3 eigenvalues are connected to visualize the anisotropy. In this case, the molecular polarizabilities are isotropic, shown by the horizontal lines between eigenvalues. The same results for the electrostatically interacting dipole model ($U_\text{exch-ind}=0$ in Equation \ref{pff}) are shown in Figure \ref{molecular_pol}(b). Compared to the non-interacting model, the root mean square deviation (RMSD) on the eigenvalues is decreased from 13.8 a.u.\ to 9.36 a.u.\ due to an improved description of the anisotropy. Lastly, including the exchange interaction ($U_\text{exch-ind}=8.13$ in Equation \ref{pff}) further improves the predicted eigenvalues to an RMSD of 7.54 a.u. The remaining error originates mainly from a systematic underestimation of the polarizability. The anisotropy of the molecular polarizabilities is also improved over those in the electrostatic model. This indicates that the exchange interaction can indeed be a useful addition to a polarizable force field model, although this will be investigated further in the following Section. More importantly, it demonstrates that the dipole response of a molecule to an external field can be adequately modeled, even by use of isotropic atomic polarizabilities.

\subsection{Benchmarking of three-body energies}
\label{threebody}

\subsubsection{Water three-body energies}

	\begin{figure}[t]
		\centering
		\includegraphics[width=\textwidth]{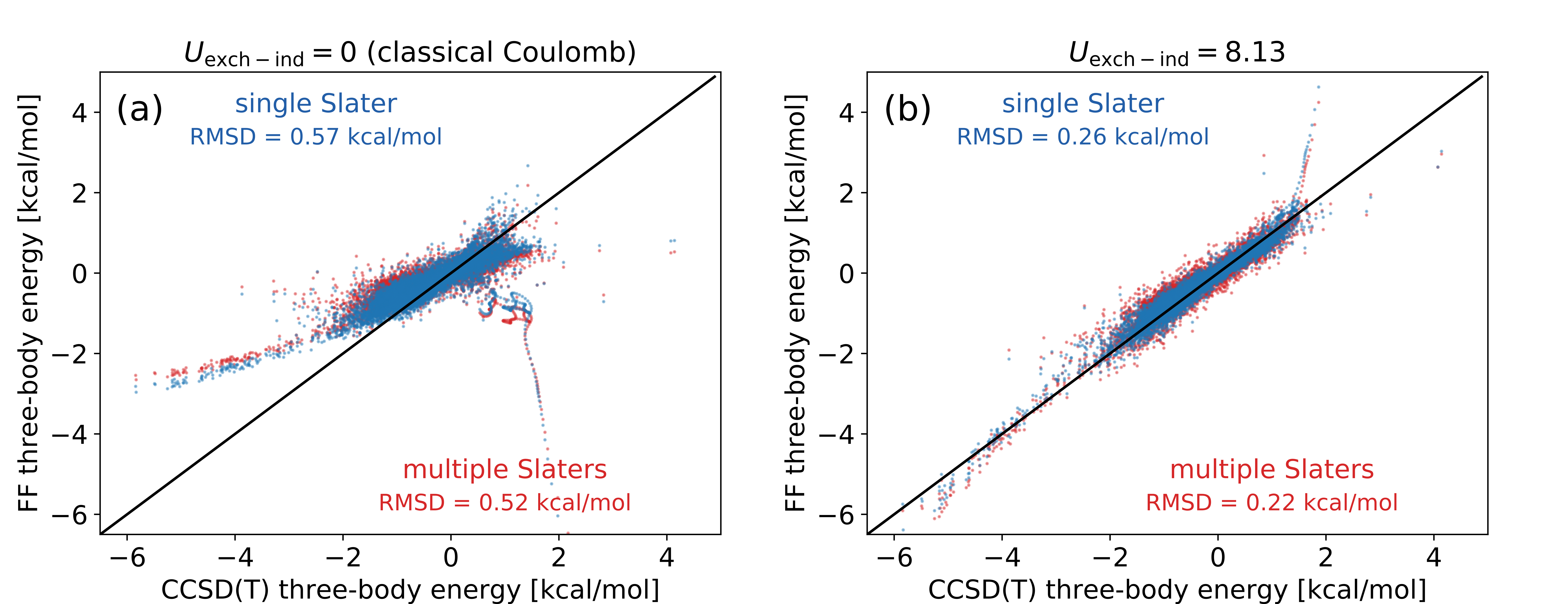}
		\caption{Parity plot of water trimer three-body energies calculated with (a) $U_\text{exch-ind}=0$ representing the classical Coulomb interaction and (b) $U_\text{exch-ind}=8.13$, computed with the single Slater fit (blue) and multiple Slater fit (red) for O and H, shown in Figure \ref{slaterwidths}.}
		\label{waterthreebody_comp}
	\end{figure}

	An important second validation of the polarizable force field is its accuracy in predicting interaction energies. In order to test the induction component separately from other intermolecular interactions, we benchmark our model on a dataset containing the three-body energies of 12347 water trimers in a wide range of conformations, calculated at the CCSD(T)/aug-cc-pVTZ level of theory.\cite{Babin2014} The total three-body energy in these systems is dominated by induction, allowing for a direct comparison with \textit{ab-initio} three-body energies.\cite{Milet1999,Podeszwa2007} First, we demonstrate the effect of including the exchange interaction in our model, using a single Slater dipole on each atom. Shown in blue in Figure \ref{waterthreebody_comp}(a) are the predicted three-body energies without the exchange interaction ($U_\text{exch-ind}=0$ in Eqs. \ref{field} and \ref{pff}), and with exchange ($U_\text{exch-ind}=8.13$) in Figure \ref{waterthreebody_comp}(b). Neglecting the exchange interaction does not yield accurate results, as the three-body energy is severely underestimated for the low-lying trimers. This trend is reversed for trimers with positive three-body energies, erroneously predicting stabilizing interactions. The inclusion of exchange results in a much improved prediction, reducing the RMSD from 0.57 kcal/mol to 0.26 kcal/mol. The same results using the multiple Slater fit is shown in red in both Figures. Only slight differences are observed compared to the single Slater fit, suggesting that the inclusion of the precise density fluctuations near the nucleus is not vital to obtain accurate three-body induction energies, and the single Slater fit can be used for improved computational efficiency. The multiple Slater fit however slightly decreases the error and will therefore be used for the all calculations in the remainder of this work.

It should be noted that the model was never fitted to three-body energies, and the exchange parameter $U_\text{exch-ind}$ was only fitted to dispersion dominated dimer interaction energies of the S66x8 dataset, which do not include water. Only the ground state density of a water monomer in the optimized geometry was used as input. The good performance across the range of monomer conformations also demonstrates our model's robustness with respect to conformational changes.

\begin{figure}[t]
	\centering
	\includegraphics[width=0.58\textwidth]{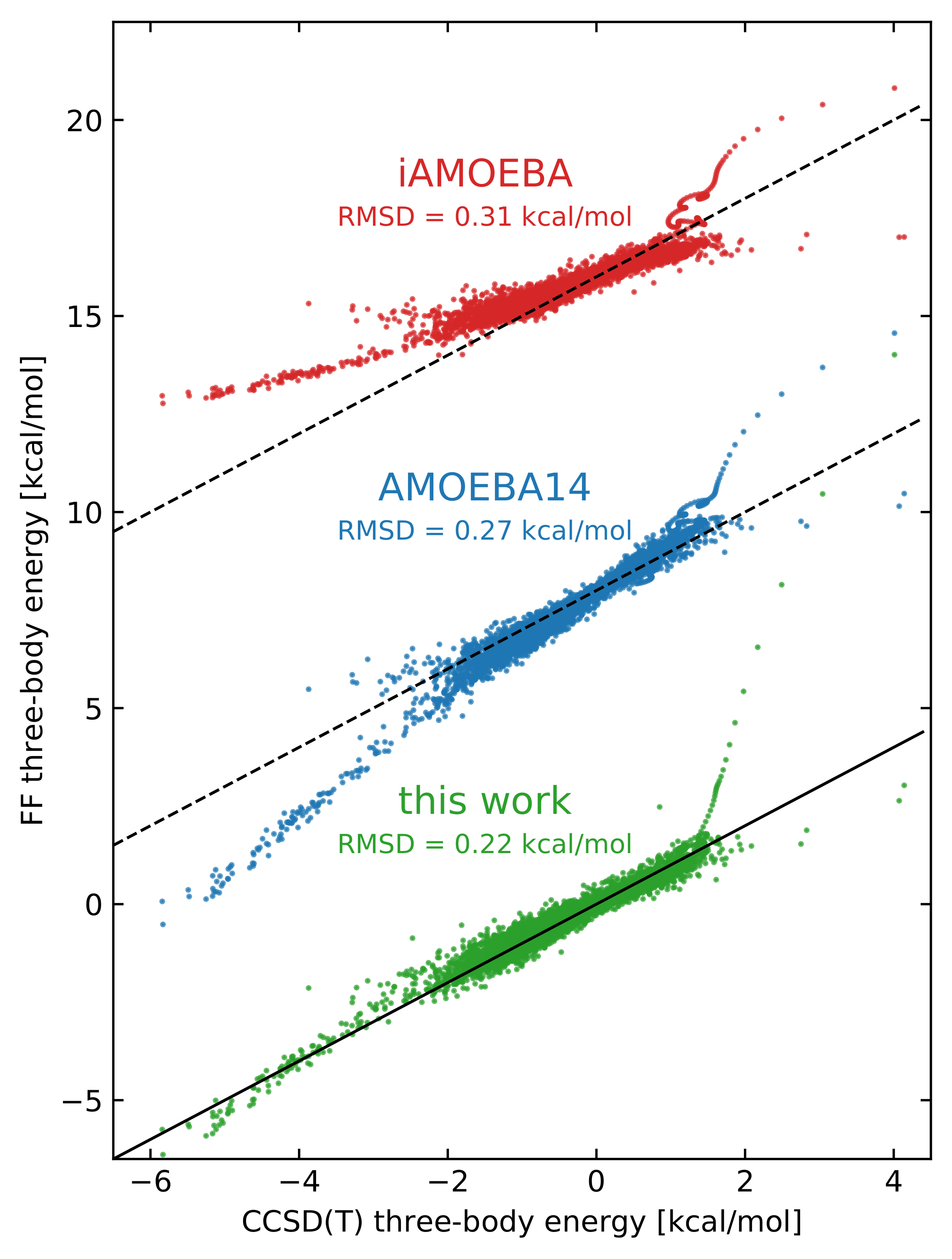}
	\caption{Parity plot of water trimer three-body energies calculated with the polarization model developed in this work (green), AMOEBA14\cite{Laury2015} (blue) and iAMOEBA\cite{Wang2013} (red). The results obtained with AMOEBA14 and iAMOEBA are shifted upward for clarity.}
	\label{waterthreebody}
\end{figure}

This trimer dataset was previously constructed to fit a full-dimensional potential energy function for water (MB-pol\cite{Babin2014}). As our polarization model aims at being transferable without fitting to the interaction energies of specific systems at hand, we cannot expect the model to be competitive on the specific case of water with MB-pol. In contrast, we compared our model with the popular iAMOEBA\cite{Wang2013} and AMOEBA14\cite{Laury2015} force fields from the literature. The results are shown in Figure \ref{waterthreebody}. AMOEBA14 tends to overestimate the magnitude of the three-body energies, an observation which has been made before.\cite{Christie2005,Kumar2010} iAMOEBA, on the other hand, underestimates the magnitude of three-body interaction due to neglecting back-polarization. The total RMSD of our model is slightly lower, at 0.22 kcal/mol, compared with 0.27 kcal/mol and 0.31 kcal/mol. This is an encouraging sign, as the AMOEBA parameters are fitted specifically to reproduce water interaction energies and condensed phase properties, while no such fitting to water was performed for our model. Especially the lower energy trimers are described well by our model, compared with both AMOEBA14 and iAMOEBA, although some large errors are seen for the high energy trimers. Inspection of these trimers revealed that these represent geometries containing unrealistically small hydrogen bond lengths of less than 1.3 \AA, rendering those less physically relevant. Removing these trimers results in a much improved RMSD of our model of 0.17 kcal/mol.

The main reason why AMOEBA14 performs well without including any exchange interaction is due to its use of point dipoles. These result in stronger interactions than delocalized dipoles, compensating for the missing exchange interaction. Note that no three-body energy appears for the non-polarizable force fields due to their pairwise additivity, resulting in an RMSD of 0.80 kcal/mol.

	\subsubsection{3b69 trimer three-body energies}

	\begin{figure}[t]
		\centering
		\includegraphics[width=0.6\textwidth]{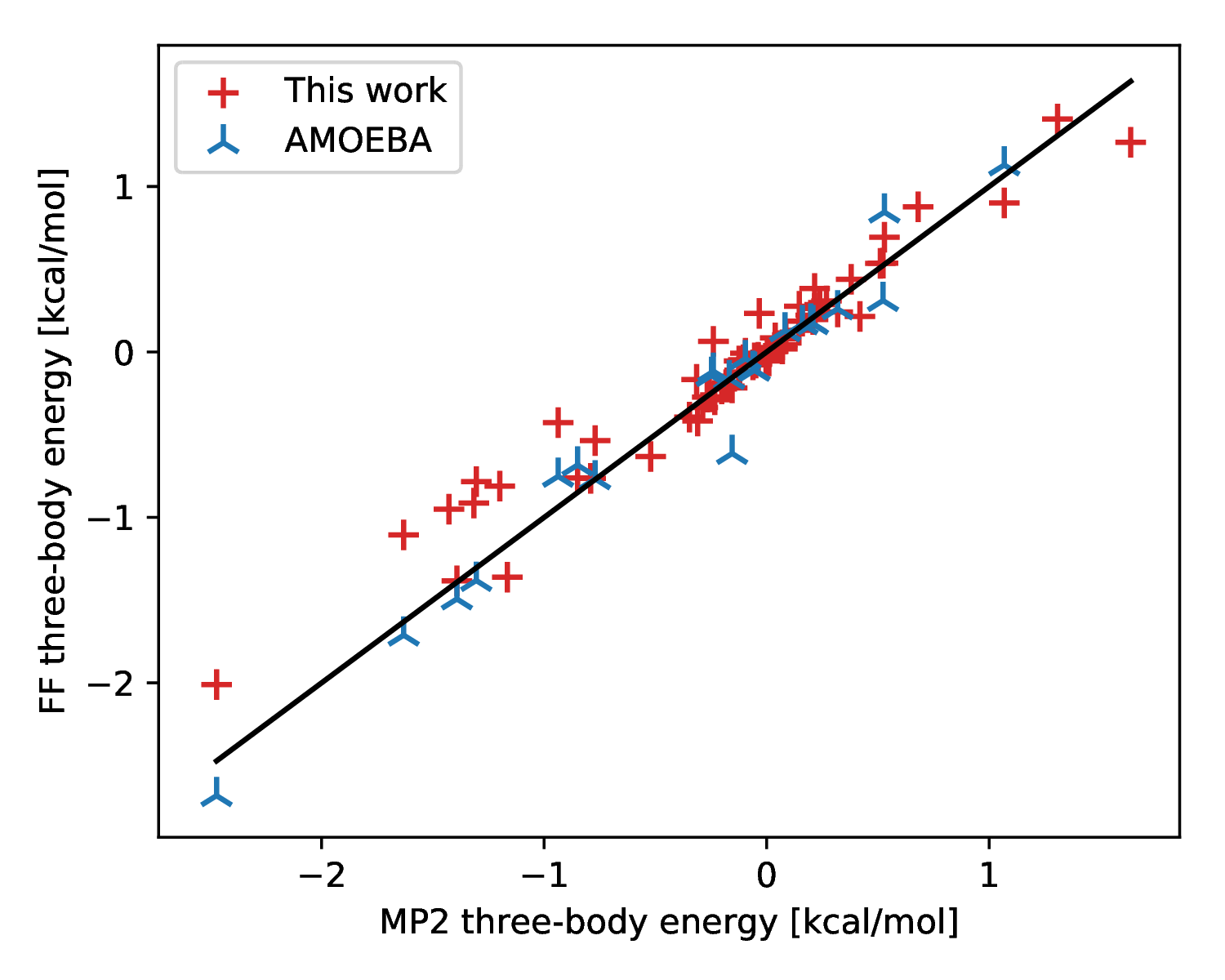}
		\caption{Parity plot of three-body energies of trimers in the 3b69 set calculated at the MP2/CBS level of theory, compared with the polarization model proposed in this work (red) and the AMOEBA force field (blue).\cite{Ren2011}}
		\label{3b69}
	\end{figure}

	 To investigate whether the polarization model performs well across a wide range of intermolecular interactions, we performed a benchmark on three-body energies of trimers in the 3b69 set.\cite{Rezac2015} This set comprises trimers with a mixture of many-body polarization and dispersion interactions. Therefore, to test the polarization component separately from dispersion, three-body energies calculated at the MP2/CBS level of theory were used as benchmark. MP2 includes many-body induction effects, but the dispersion non-additivity only appears at the MP3 level \cite{Rezac2015, Chalasinski1994}, allowing for a direct comparison between MP2 three-body energies and those predicted by the inducible dipole model. Again, only the B3LYP/aug-cc-pVTZ ground state monomer electron densities of the molecules present in the dataset were used as input. The predicted three-body energies (shown in red in Figure \ref{3b69}) result in a RMSD of 0.20 kcal/mol, demonstrating a good performance across the range of interactions. For comparison, we calculated the same three-body energies with the AMOEBA force field\cite{Ren2011} for the 22 out of 69 trimers for which parameters were available. For this subset, the RMSD on the three-body energies is 0.16 kcal/mol. Our model's comparable performance with AMOEBA is very encouraging, as 18 out of the 24 molecules present in the 3b69 set are not present in the S66x8 set, to which the proportionality factor between the overlap in electron density and the exchange-repulsion energy $U_\text{exch-ind}$ was fitted. This suggests that the model transfers well to molecules outside the training set.

	\subsection{Development of a full non-covalent force field}
	\label{p-medff}

	Until now, we have only compared the polarization component to \textit{ab-initio} three-body energies. To predict total non-covalent interaction energies, we incorporate our new polarization model in the recently developed monomer electron density force field (MEDFF). In MEDFF, interaction energies are decomposed in four terms in accordance with SAPT. \cite{Vandenbrande2017} Of particular interest are the exchange-repulsion and induction terms, both of which were represented as a proportionality between the overlap integral of electron densities:

	\begin{align}
		E_\text{exch-rep}&=U_\text{exch-rep}\sum_i^{N_1}\sum_j^{N_2}S_{ij}\quad \mathrm{with}\quad S_{ij}=\int\rho^{1s}_i (\vec{r})\rho^{1s}_j (\vec{r})\d\vec{r}\label{exch-rep}\\
		E_\text{ind}&=-U_\text{ind}\sum_i^{N_1}\sum_j^{N_2}S_{ij}
	\end{align}

	\begin{figure}[t]
		\centering
		\includegraphics[width=0.6\textwidth]{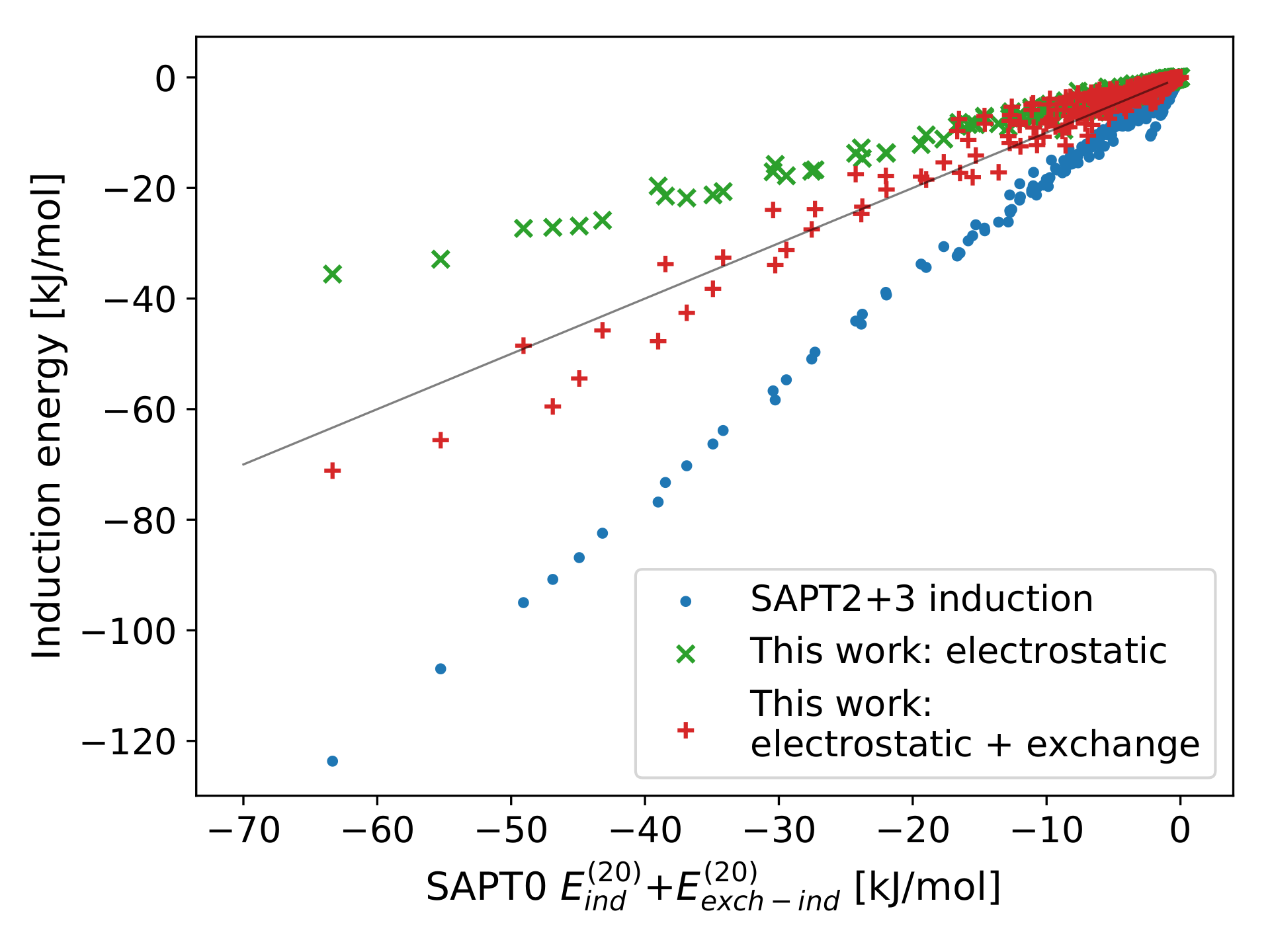}
		\caption{Parity plot of dimer induction energies of the S66x8 set computed with SAPT2+3 and our polarization model with and without the exchange interaction, compared to the SAPT0 $E^{(20)}_\text{ind}+E^{(20)}_\text{exch-ind}$ energy.}
		\label{ind}
	\end{figure}

	The sums over $i$ and $j$ run over all atoms of the first and second monomer, respectively. Our polarization model now replaces the induction term. However, we cannot expect a physical polarization model to reproduce the full SAPT2+3 induction interaction, as it includes higher-order terms, both in the intermolecular and intramolecular order (although higher order intramolecular terms could be captured in the atomic polarizabilities). \cite{Parker2014} In the lowest order, the SAPT0 induction term can be written as follows:

	\begin{equation}
		E_\text{ind}^\text{SAPT0}=E_\text{ind}^{(20)}+E_\text{exch-ind}^{(20)}+\delta E_\text{HF}^{(2)}
	\end{equation}

	Where the first and second superscript denote the intermolecular and intramolecular order, respectively. $\delta E_\text{HF}^{(2)}$ is a term representing polarization beyond the second order. \cite{Parker2014} We can therefore only expect our polarization model to reproduce the first two terms of $E_\text{ind}^\text{SAPT0}$. This is corroborated by a direct comparison between our model and the first two SAPT0 terms for the dimers in the S66x8 set, made in Figure \ref{ind}. In higher orders of SAPT, the exchange and induction components mix and can no longer be separated in a pure exchange and induction component.\cite{Parker2014} Inspired by this fact, $U_\text{exch-rep}$ can be refit to incorporate the missing higher order induction terms beyond the first two terms in $E_\text{ind}^\text{SAPT0}$. In this way, we accurately capture the many-body component of induction (as evidenced by the benchmarking on three-body energies), as well as the two-body interaction (by adding the missing interaction terms to $U_\text{exch-rep}$). In summary, our complete non-covalent force field, termed the polarizable monomer electron density force field (PMEDFF) is as follows:

	\begin{equation}
		E\Big[U_\text{exch-rep}, U_\text{exch-ind}, U_\text{s8}\Big]=E_\text{elst}+E_\text{exch-rep}\Big[U_\text{exch-rep}\Big]+E_\text{disp}\Big[U_\text{s8}\Big]+E_\text{ind}\Big[U_\text{exch-ind}\Big]
	\end{equation}

	where the electrostatic and dispersion terms are the same as in MEDFF \cite{Vandenbrande2017}, the exchange-repulsion term is given by Eq.\ \ref{exch-rep}, and the induction term is described in Section \ref{interaction}. The proportionality between the overlap and the exchange-repulsion interaction, $U_\text{exch-ind}$, was previously fit to the S66x8 set and was set to 8.13 a.u. \cite{Vandenbrande2017} Finally, refitting $U_\text{exch-rep}$ to the SAPT2+3 exchange-repulsion and higher order induction terms for the whole S66x8 set yields a value of 6.64 a.u. However, this parameter remains somewhat sensitive to the type of intermolecular interaction. Refitting to only the electrostatically or dispersion dominated dimers in the S66x8 set yield values of 6.15 a.u. and 8.18 a.u., respectively. Therefore it is useful to refit $U_\text{exch-rep}$ to the system or interaction type of interest. The other parameters will remain fixed in the remainder of this work. With the new force field constructed, we now validate it on interaction energies of dimers not included in the S66x8 set.

	\subsection{Benchmarking of dimer interaction energies}
	\label{dimer}

	\begin{figure}[t]
		\centering
		\includegraphics[width=\textwidth]{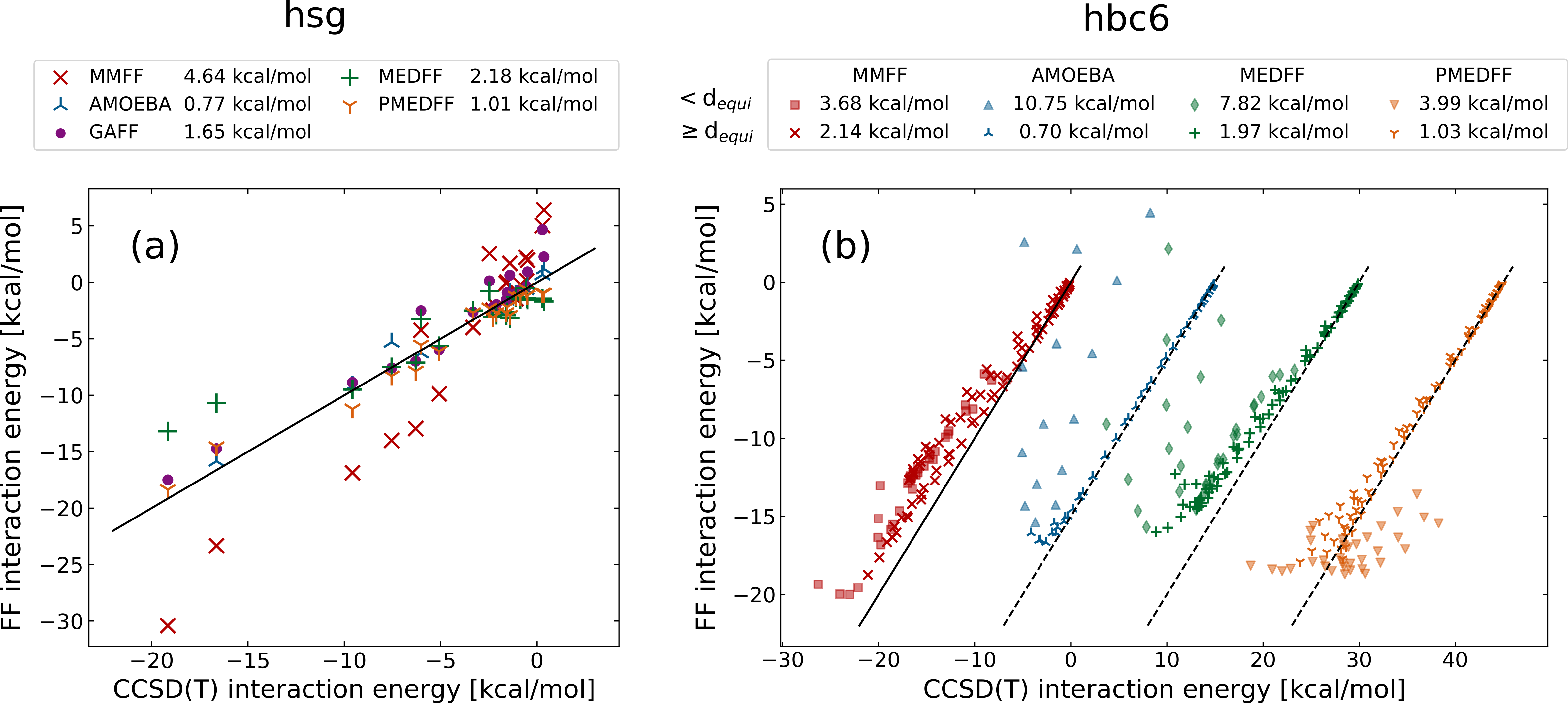}
		\caption{(a) Parity plot of dimer interaction energies of the hsg set calculated with MMFF\cite{Halgren1999} (red), AMOEBA\cite{Ren2011} (blue), GAFF\cite{Wang2004} (purple), MEDFF\cite{Vandenbrande2017} (green), and PMEDFF with $U_\text{exch-rep}=6.64$ a.u.\ (orange), and (b) the hbc6 set calculated with MMFF (red), AMOEBA (blue), MEDFF (green) and PMEDFF with $U_\text{exch-rep}=6.15$ a.u.\ (orange). The RMSD for each force field is shown in the legend. For the hbc6 set, dimers along the dissociation curve with rescaled intermolecular distances smaller than the equilibrium distance ${d_\text{equi}}$ are displayed with filled symbols in a lighter shade. Results of the hbc6 set for all force fields except MMFF are shifted horizontally for clarity.}
		\label{hsg-hbc}
	\end{figure}

	To validate our force field, we test its performance on two datasets of dimers which are not present in the S66x8 set to which the interaction parameters were fit. Both datasets and the associated interaction energies computed at the CCSD(T)/CBS level of theory were taken from the BioFragment Database.\cite{Burns2017,Burns2014,Faver2011} The hsg set contains 21 dimer fragments, extracted from an HIV-II protease crystal structure with a bound ligand (indinavir), representing a wide range of interactions (from dispersion to electrostatically dominated).\cite{Faver2011} Therefore, the universal value for $U_\text{exch-rep}$ of 6.64 a.u.\ was used for PMEDFF. A comparison was made with the Merck molecular force field\cite{Halgren1999} (MMFF), AMOEBA\cite{Ren2011}, the generalized amber force field\cite{Wang2004} (GAFF) and MEDFF\cite{Vandenbrande2017}. As can be seen on the left of Figure \ref{hsg-hbc}, PMEDFF performs well on the whole set, with an RMSD of 1.01 kcal/mol. Both GAFF and AMOEBA also show a balanced performance across the range of interaction types, although the RMSD of 0.77 kcal/mol obtained for AMOEBA only includes 13 out of the 21 complexes for which parameters are available. A significant overestimation of the magnitude of the interaction energy of the electrostatically dominated complexes is observed for MMFF, resulting in an RMSD of 4.64 kcal/mol. MEDFF, on the other hand, slightly underestimates the interaction energy of some of the electrostatically dominated complexes, yielding an RMSD of 2.18 kcal/mol.

	The hbc6 set consists of doubly hydrogen bonded dimers extracted from 6 dissociation curves containing formic acid, formamide and formamidine.\cite{Thanthiriwatte2011} Because of the interaction type, the $U_\text{exch-rep}$ parameter was set to 6.15 a.u., obtained from the fit of only electrostatically dominated complexes of the S66x8 set (which do not contain formic acid, formamide or formamidine).\cite{Rezac2011} Due to the double hydrogen bonds, the induction component of the interaction for complexes around the equilibrium intermolecular distances is much larger than for dimers in the hsg set.\cite{Burns2017} This set is therefore a more stringent test of the induction component of our force field. We compare with AMOEBA, MMFF and MEDFF. Only 60 out of the 118 dimers containing formic acid and formamide were retained for AMOEBA, as no parameters were available for formamidine. We divide our comparisons with other force fields between dimers with an intermolecular distance larger than and smaller than that of the optimized dimer geometry. For large intermolecular distances, the RMSD of MMFF, AMOEBA, MEDFF and PMEDFF is equal to 2.14 kcal/mol, 0.70 kcal/mol, 2.27 kcal/mol and 1.03 kcal/mol, respectively. The dimers with rescaled intermolecular distances smaller than 1 are more challenging to predict, as hydrogen bonds are artificially compressed. The increase in error from the rescaled distances larger than 1 is most pronounced for AMOEBA. The error increases to 10.75 kcal/mol, due to the prediction of large repulsive interaction energies. Lower errors of 5.77 kcal/mol, 3.68 kcal/mol and 3.99 kcal/mol are seen for MEDFF, MMFF and PMEDFF. Overall, the performance of PMEDFF against other force fields is encouraging, especially given that, as a test of transferability of our force field, no fitting was performed to any of the dimers present in both the hsg and hbc6 set.

	\subsection{Many-body induction in the condensed phase of water}
	\label{liquidwater}

\begin{figure}[t]
	\centering
	\includegraphics[width=\textwidth]{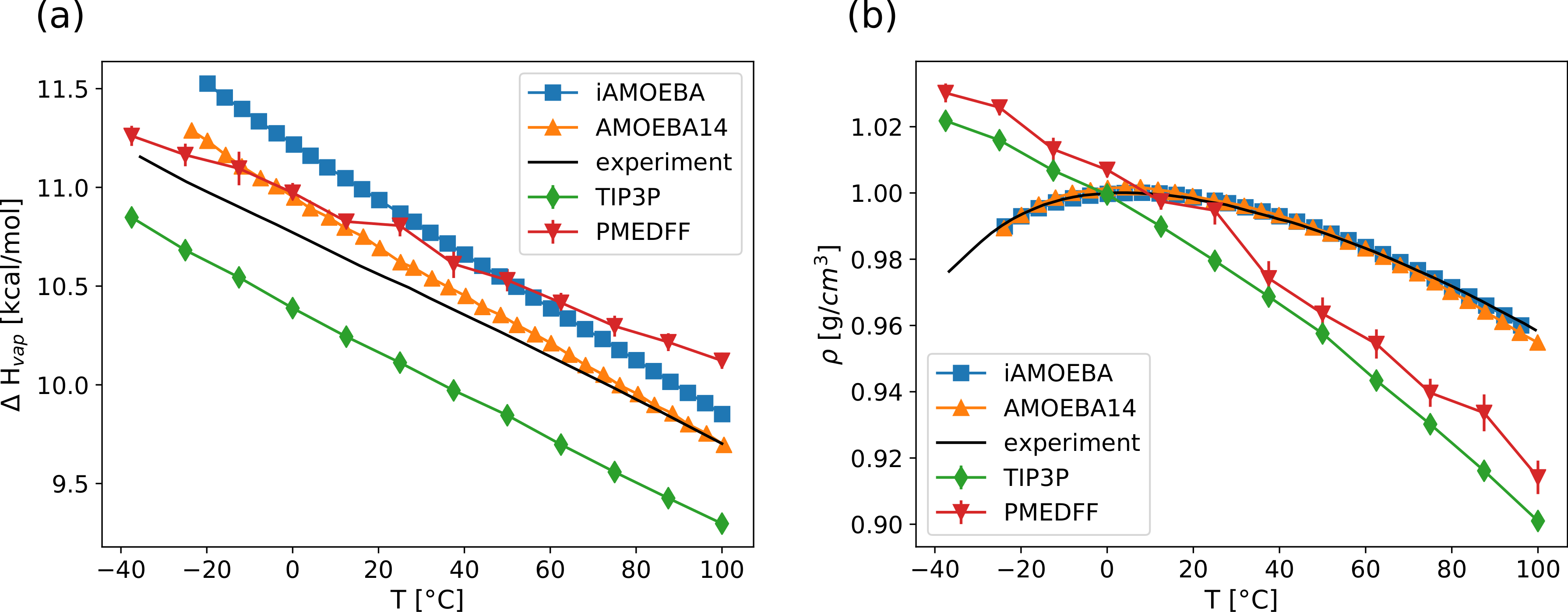}
	\caption{(a) Heat of vaporization and (b) density of water obtained with iAMOEBA\cite{Wang2013}, AMOEBA14\cite{Laury2015}, TIP3P\cite{Price2004} and PMEDFF compared with experiment over a temperature range of $-40\;^\circ \mathrm{C}$ to $100\;^\circ \mathrm{C}$. The 1$\sigma$ uncertainty intervals for TIP3P and PMEDFF were obtained from 3 independent simulations.}
	\label{Hwater}
	\end{figure}

	Previously, we showed that three-body energies of water are predicted well by PMEDFF. To verify whether this performance is maintained in the condensed phase, we calculated the heat of vaporization $\Delta H_\text{vap}\approx k_\text{B}T-E_\text{pot}$ of water from rigid-body NPT Monte Carlo simulations on a box containing 150 molecules\cite{Caleman2012,10.5555/559571}. The exchange-repulsion interaction parameter $U_\text{exch-rep}$ was refit solely on the dissociation profile of 8 water dimers contained in the S66x8 set, yielding a value of 7.45 a.u. An initialization run of $2\times10^6$ MC steps was followed by a production run of $8\times10^6$ steps. The resulting heats of vaporization over a temperature range of $-40\;^\circ \mathrm{C}$ to $100\;^\circ \mathrm{C}$ are shown in Figure \ref{Hwater}(a), and compared with iAMOEBA, AMOEBA14 and TIP3P. The iAMOEBA and AMOEBA14 results were obtained from the literature.\cite{Wang2013,Laury2015} For TIP3P, NPT MD simulations were performed in the Tinker program.\cite{Rackers2018,10.5555/559571} The long-range electrostatics was calculated with particle mesh Ewald (PME), and van der Waals interactions were cut off at a distance of 8 \AA and supplemented with analytical tail corrections. An Andersen thermostat and Berendsen barostat were used with the default coupling constants present in Tinker. Intermolecular geometries were constrained with SHAKE.

	The heats of vaporization predicted by PMEDFF are in good agreement with experiments, comparable with the polarizable iAMOEBA and AMOEBA14 water models, demonstrating that our force field transfers well from the gas phase to the condensed phase. It should be noted that, as our model is derived from the gas phase, it does not implicitly capture nuclear quantum effects. The hypothetical heat of vaporization for classical water without these quantum effects at 298.15 K has been calculated to be 11.0 kcal/mol.\cite{Guillot2001} This value is in excellent agreement with the heat of vaporization of 10.8 kcal/mol calculated by PMEDFF. The temperature dependence of the density is not predicted well by PMEDFF and is comparable with TIP3P. Here the iAMOEBA and AMOEBA14 models are clearly superior. This can probably be attributed to the better description of the permanent atomic multipoles in iAMOEBA and AMOEBA14, as only monopoles are used in PMEDFF and TIP3P. However, the density predicted by PMEDFF at 298.15 K is in excellent agreement with experiments. This is a fortuitous coincidence, as the  $U_\text{exch-rep}$ parameter was solely fitted to 8 gas phase dimer interactions, and not to the condensed phase at this temperature.

	\subsection{Guest adsorption of CO$_2$ and H$_2$O in metal-organic framework ZIF-8}
	\label{mof}

	The main advantage of the model developed in this work is its transferability, both to the condensed phase, as shown in the previous Section, but also to periodic structures. We demonstrate this by applying our model to the prediction of CO$_2$ and H$_2$O insertion energies in the metal-organic framework (MOF) ZIF-8. The all-electron density of the framework needed as input for our force field was obtained from a periodic PBE\cite{Perdew1996,Grimme2011} calculation with a cutoff of 600 eV, performed in GPAW.\cite{Enkovaara2010,Mortensen2005} Free atom densities were calculated with PBE in the same basis sets as in Section \ref{widths}, including the X2C relativistic correction for Zn. The \textit{ab-initio} reference energies were obtained with VASP\cite{Kresse1993,Kresse1994,Kresse1996,Kresse1996a} at the PBE+D3(BJ) level of theory using the projector augmented wave (PAW) method.\cite{Blochl1994,Joubert1999} A fully converged \textit{ab-initio} calculation of the adsorption energy at infinite dilution, given by

	\begin{equation}
	E_\text{ads}=\frac{\int \Delta U e^{-\beta \Delta U} \d\vec{s}}{\int e^{-\beta \Delta U} \d\vec{s}}
	\end{equation}

	\begin{figure}[t]
		\centering
		\includegraphics[width=0.45\textwidth]{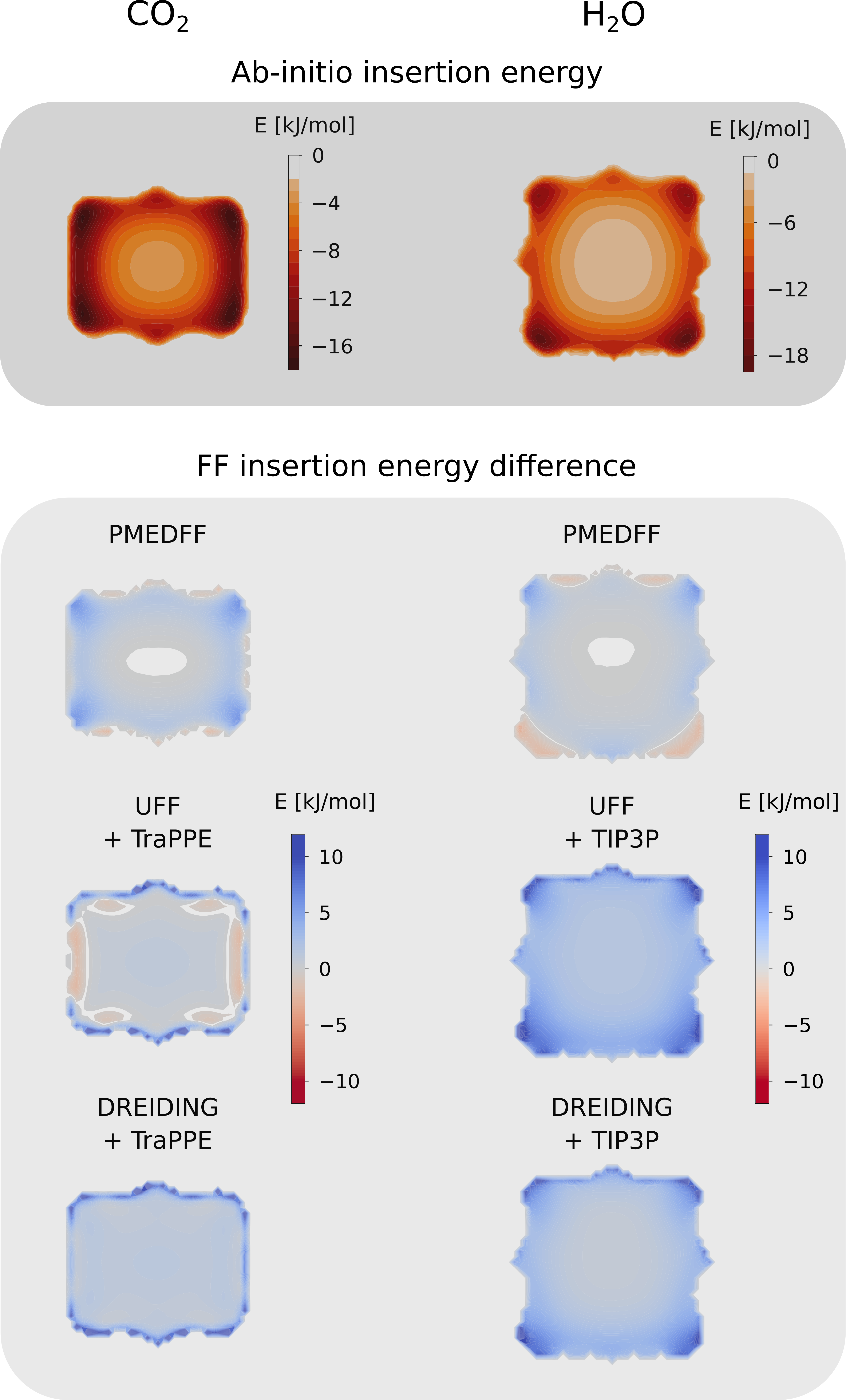}
		\caption{ (top) \textit{Ab-initio} insertion energies of CO$_2$ and H$_2$O. (bottom) Difference between the force field and \textit{ab-initio} insertion energies predicted by PMEDFF, UFF\cite{Rappe1992}, and DREIDING\cite{Mayo1990}. The UFF and DREIDING force field are supplemented with TraPPE\cite{Chen1999} for CO$_2$ and TIP3P\cite{Price2004} for H$_2$O.}
		\label{zif}
	\end{figure}

	with $\Delta U$ the insertion energy and $\beta=\frac{1}{k_\text{B}T}$, would require on the order of 10$^6$ insertions, making it computationally extremely demanding and only feasible for small and highly symmetric unit cells.\cite{Vandenbrande2018} Therefore, we calculated insertion energies on a grid with a density of 16 \AA$^{-2}$ in the yz-plane through the center of the ZIF-8 unit cell. A fixed y-aligned orientation of the adsorbates was chosen, as a rotational scan at each point is too computationally demanding. Insertions with positive \textit{ab-initio} energies were discarded, leaving a total of 852 insertions for CO$_2$ and 1071 for H$_2$O. ZIF-8 is composed of tetrahedrally coordinated zinc ions connected by imidazolate linkers. Therefore, we fitted the $U_\text{exch-rep}$ parameter to 10 points along the dissociation curve of imidazole and both CO$_2$ and H$_2$O, yielding values of 7.22 a.u.\ and 7.43 a.u., respectively. The optimization was performed in Psi4\cite{Parrish2017} at the MP2/cc-pVTZ level of theory, after which the distance between the center of masses of the monomers was rescaled by between 0.8 and 2 times the equilibrium distance. The final interaction energies were calculated at the CCSD(T)/CBS level of theory by extrapolation of the MP2 energy in the aug-cc-pVTZ and aug-cc-pVQZ basis sets using the Helgaker scheme\cite{Halkier1998}, and a calculation of the CCSD(T) correction in the aug-cc-pVDZ basis set. This method was previously used in the construction of the S66x8 set.\cite{Ren2011}. The performance of PMEDFF is compared with UFF and the Lennard-Jones potential of the DREIDING force field combined with TraPPE for CO$_2$ and TIP3P for H$_2$O. \cite{Rappe1992,Mayo1990,Chen1999,Price2004} UFF is supplemented with charges obtained with the extended charge equilibration (EQeq) method\cite{Wilmer2012}, while no charges were assigned for the DREIDING force field. Differences between the \textit{ab-initio} and force field insertion energies are shown in Figure \ref{zif}. The RMSD of PMEDFF is equal to 1.92 kJ/mol and 2.23 kJ/mol for CO$_2$ and H$_2$O, compared with 3.04 kJ/mol and 7.45 kJ/mol for UFF and 3.66 kJ/mol and 6.25 kJ/mol for DREIDING. As can be seen in Figure \ref{zif}, the error observed for UFF and DREIDING originates mainly from an overly repulsive interaction close to the framework, caused both by the lack of an induction term which stabilizes the hydrogen bonding interaction with the framework, as well as the differing functional form of exchange-repulsion (exponential form of PMEDFF compared to the 12-6 Lennard-Jones potential). Additionally, both UFF and DREIDING consistently underestimate the interaction energy in the center of the ZIF-8 pore, while this error is much less pronounced for PMEDFF. This suggests that the electrostatic and dispersion functionals also perform well in their transferability, as they were not fitted to the system at hand.

	\section{Conclusions and outlook}

	A new transferable polarization model was developed, based on the often-used inducible dipole model. It is transferable in the sense that only the ground state electron density of the molecule or periodic structure in question is required, together with a single proportionality factor which was fitted to the S66x8 set. Slater dipoles were introduced as the dipole response functions, and exchange-repulsion was included as an additional interaction inducing the Slater dipoles. The resulting model performs well on the prediction of three-body energies of trimers containing a wide range of intermolecular interactions. Importantly, this is also the case for molecules not present in the S66x8 set and molecules for which no specifically fitted parameters are available for induction models in the literature.

	A complete polarizable non-covalent force field, coined the polarizable monomer electron density force field (PMEDFF), was developed by including our polarization model in the previously proposed monomer electron density force field (MEDFF). A benchmark on two dimer datasets revealed a performance comparable or better than force fields in the literature, without the need for molecule specific parameters. Moreover, the inclusion of many-body induction in our force field results in an accurate prediction of the heat of adsorption of water in the condensed phase. We concluded by presenting a possible use case of our force field; guest adsorption in metal-organic frameworks. For these materials, specifically fitted polarization models are usually not available, while our force field can nevertheless be applied. A significant improvement of the predicted insertion energies was observed, compared to transferable force fields in the literature.

	Our new polarization model shows clear promise in improving the description of hydrogen-bonded structures, both in the gas phase and condensed phase. However,
	PMEDFF still uses a relatively simple isotropic model for the atomic polarizabilities, resulting in noticeable errors in the description of the molecular anisotropy. A more accurate determination of the atomic response is therefore an important avenue for future work. Additionally, as our force field was not fitted to experimental condensed phase properties, no nuclear quantum effects (NQE) are implicitly included. The explicit inclusion of these effects could therefore further increase the accuracy of our force field in predicting condensed phase properties. Moreover, our model retains a pair-wise additive dispersion model, neglecting any many-body dispersion interactions. These effects certainly become important in the condensed phase, and future extensions of our force field will focus on including an appropriate model to include these interactions.

	\section*{Author information}

	\subsection*{Corresponding Author}
	*E-mail; Toon.Verstraelen@UGent.be

	\subsection*{Notes}
	The authors declare no competing financial interest.

	\section*{Acknowledgement}
	This research was funded by the Research Board of Ghent University (BOF). The computational resources and services used were provided by Ghent University (Stevin Supercomputer Infrastructure).

	\section*{Supporting Information Avaiable}
	The fitting procedure and results for the Slater dipole widths and amplitudes.

	\bibliography{bib.bib}

\end{document}


\maketitle
	
	
	\vfill
	
	\textbf{Keywords:} inducible dipole model, polarizable force field, non-covalent force field
	
	\vspace{1cm}
	
	\newpage
	
	\section{Fitting of slater widths}
	Perturbed and unperturbed free atom densities were obtained at the CCSD/aug-cc-pV5Z \cite{Purvis1982,Peterson1994} level of theory using Gaussian09\cite{g09} for H, C, N, O, F, Mg, Al, P, S and Cl. For Zn, the exact-two-component (X2C) method\cite{Verma2016} at the CCSD/ANO-RCC-VQZP level of theory\cite{Pritchard2019} was used in Psi4\cite{Parrish2017} to include scalar relativistic effects. The perturbation consists of a dipole field with a strength of 0.0001 a.u in the $z$-direction. To fit a unit dipole (composed of one or multiple Slater functions), the \textit{ab-initio} density difference $\rho^{\text{diff}}$ between the perturbed and unperturbed densities is computed on a Becke-Lebedev grid \cite{Becke1988,LebedevV.I.;Laikov1999}, using the open-source code HORTON.\cite{Verstraelen2015} The density difference is normalized by computing the expectation value of $z$ on this grid and dividing the density difference by $\langle z_{\text{diff}}\rangle$:
	
	\begin{equation}
	\langle z_{\text{diff}}\rangle=\int\rho^{\text{diff}}(\vec{r})z \d\vec{r}
	\end{equation}
	
	\begin{equation}
	\rho^{\text{diff}}_{\text{norm}}(\vec{r}) = \frac{\rho^{\text{diff}}(\vec{r})}{\langle z_{\text{diff}}\rangle}
	\end{equation}

	\subsection{Single Slater fit}
	
	The normalized density difference is now used to fit a single normalized Slater dipole along the $z$-axis of the following form:
	
	\begin{equation}
	\rho_z(\vec{r})=\frac{S^5}{32\pi}\exp(-Sr)\,r\cos\theta
	\end{equation}
	
	A straightforward least squares fit of the exponent $S$ would yield unreliable results, as large fluctuations of the density difference occur near the nucleus. Therefore, we opt to fit the third moment of $z$ of the normalized Slater dipole $\langle z_{\text{Slater}}^3\rangle$ to that of the normalized density difference $\langle z_{\text{diff}}^3\rangle$. The exponent $S$ is thus determined as follows:
	
	\begin{equation}
	\langle z^3_{\text{diff}}\rangle = \langle z^3_{\text{Slater}}\rangle=\int\rho_z(\vec{r})z^3 \d\vec{r}=\frac{18}{S^2}\qquad \Longrightarrow \qquad S=\sqrt{\frac{18}{\langle z^3_{\text{diff}}\rangle}}
	\end{equation}
	
	The resulting Slater exponents are shown in Table \ref{widths}.
	
	\subsection{Multiple Slater fit}
	
	To investigate the effect of the large density fluctuations near the nucleus on the polarization energy, we also performed a fitting of multiple Slater functions, capturing the density fluctuations. This was done by computing the normalized density difference on a 1D grid along the $z$-axis. For every atom, the amplitudes and exponents of $N$ Slater functions were fit to the density difference on this grid with a least squares regression. The sum of $N$ amplitudes was constrained to 1, to obtain a normalized multiple Slater dipole. The number of Slater functions $N$ was varied from 1 to 6 for each element, and the lowest number able to capture the density fluctuations was kept.
	\newline
	
	\begin{table}
		\setlength\arrayrulewidth{0.5pt}
	\rowcolors{2}{gray!25}{white}
	\caption{Single Slater exponents and multiple Slater amplitude and exponents for some common elements.}
	\label{widths}
	\begin{center}
	\scalebox{0.7}{
	\begin{tabularx}{1.2\linewidth}{c | c | X}
		\rowcolor{gray!50}
		Element & \hfill Single Slater exponents [a.u.] \hfill &
		\hfill \qquad Multiple Slater amplitudes \& exponents [a.u.] \qquad \hfill\null \\
		
		H & \makecell{ \\ 1.8076} & \hfill \makecell{1.0000 \\ 1.7378} 
		\hfill\null \\
		
		C & \makecell{ \\ 1.5476} & \hfill \makecell{0.0259 \hspace{0.2cm} 14.704 \hspace{0.2cm} -13.729 \\ 4.5397 \hspace{0.2cm} 2.1112 \hspace{0.2cm}\;\, 2.1619} \hfill\null \\
		
		N & \makecell{ \\ 1.8913} & \hfill \makecell{15.058  \hspace{0.2cm} 2.6330 \hspace{0.2cm} 18.206 \hspace{0.2cm} -2.3654 \hspace{0.2cm} -32.532 \\ 3.0879 \hspace{0.2cm}  0.9323 \hspace{0.2cm}  2.8248 \hspace{0.2cm} 0.8743 \;\,\hspace{0.2cm} 2.9684} \hfill\null \\
		
		O & \makecell{ \\ 1.6744} & \hfill \makecell{-27.652\hspace{0.2cm}  0.0088\hspace{0.2cm}  28.643 \\ \;\,2.7294\hspace{0.2cm}  10.805\hspace{0.2cm}  2.6887} \hfill\null \\
		
		F & \makecell{ \\ 2.0842} & \hfill \makecell{-0.2884\hspace{0.2cm}  1.2714\hspace{0.2cm}  0.0170 \\ \;\,4.6097\hspace{0.2cm}  2.2994\hspace{0.2cm} 10.403} \hfill\null \\
		
		Mg & \makecell{ \\ 0.6929} & \hfill \makecell{38.090\hspace{0.2cm}  28.370\hspace{0.2cm}  2.762e-4\hspace{0.2cm} -0.0110\hspace{0.2cm} -65.4491 \\ 1.4743\hspace{0.2cm}  1.6591\hspace{0.2cm}\;\,\, 14.460\hspace{0.2cm} \; 5.3416\hspace{0.2cm}\;\,\,  1.5645} \hfill\null \\
		
		Al & \makecell{ \\ 0.8737} & \hfill \makecell{-30.285\hspace{0.2cm}  1.177e-4\hspace{0.2cm}  17.696\hspace{0.2cm} -0.0086\hspace{0.2cm} 13.598 \\ 1.7777\hspace{0.2cm} \;\,17.274\;\;\hspace{0.2cm}  1.6284\hspace{0.2cm}\;\, 5.5095\hspace{0.2cm}  1.9117} \hfill\null \\
		
		P & \makecell{ \\ 1.3837} & \hfill \makecell{7.702e-5\hspace{0.2cm} -5.7252\hspace{0.2cm}  5.3444\hspace{0.2cm} 1.3987\hspace{0.2cm} -0.0179 \\ \;\,20.355\hspace{0.2cm} \;\, 3.0461\hspace{0.2cm}  3.1097\hspace{0.2cm} 1.5105\hspace{0.2cm} \, 5.7617} \hfill\null \\
		
		S & \makecell{ \\ 1.3380} & \hfill \makecell{1.6584\hspace{0.2cm} -1.5945\hspace{0.2cm}  5.403e-4\hspace{0.2cm} -36.898\hspace{0.2cm}  37.834 \\ 6.2099\hspace{0.2cm} \, 6.2885\hspace{0.2cm}\; 19.819\hspace{0.2cm} \;\, 2.0810\hspace{0.2cm}  2.0558} \hfill\null \\
		
		Cl & \makecell{ \\ 1.5651} & \hfill \makecell{6.064e-4\hspace{0.2cm} -20.609\hspace{0.2cm} -2.9150\hspace{0.2cm}  2.9667\hspace{0.2cm} 21.557 \\ 20.977\;\,\;\,\hspace{0.2cm}  2.2389\hspace{0.2cm} \;\, 6.9961\hspace{0.2cm}  6.9545\hspace{0.2cm}  2.1933} \hfill\null \\

		Zn & \makecell{ \\ 1.2968} & \hfill \makecell{1.0613\hspace{0.2cm}  4.4973\hspace{0.2cm} -4.2707\hspace{0.2cm} 0.0042\hspace{0.2cm} -0.0039\hspace{0.2cm}  -0.2882 \\ 1.2474\hspace{0.2cm} 5.5405 \hspace{0.2cm} 5.4430 \hspace{0.2cm}  29.404\hspace{0.2cm} 29.960\hspace{0.2cm}\; 6.4849} \hfill\null \\

	\end{tabularx}}
	\end{center}
	\caption{Fitting results of Slater functions to the \textit{ab-initio} density difference. Exponents of the normalized single fit are shown in the second column. The results for the multiple Slater fit are shown in the third column, with the first row for each element containing the amplitudes and the second row containing the exponents.}
	\end{table}

    \FloatBarrier

    \bibliography{bib.bib}